\documentclass[a4paper]{jpconf}
\usepackage{graphicx}

\usepackage{xspace}

\usepackage{relsize}
\def\babar{\mbox{\slshape B\kern-0.1em{\smaller A}\kern-0.1em
    B\kern-0.1em{\smaller A\kern-0.2em R}}}

\def\babars{\mbox{\slshape B\kern-0.1em{\smaller A}\kern-0.1em
    B\kern-0.1em{\smaller A\kern-0.2em R\hspace{0.5em}}}}

\def\Kbar  {\kern 0.2em\overline{\kern -0.2em K}{}\xspace}

\def\Kz    {\ensuremath{K^0}\xspace}
\def\Kzb   {\ensuremath{\Kbar^0}\xspace}
\def\KzKzb {\ensuremath{\Kz \kern -0.16em \Kzb}\xspace}
\def\Kp    {\ensuremath{K^+}\xspace}
\def\Km    {\ensuremath{K^-}\xspace}

\def\KpKm  {\ensuremath{\Kp \kern -0.16em \Km}\xspace}
\def\KS    {\ensuremath{K^0_{\scriptscriptstyle S}}\xspace}

\def\Dbar    {\kern 0.2em\overline{\kern -0.2em D}{}\xspace}

\def\Dz      {\ensuremath{D^0}\xspace}
\def\Dzb     {\ensuremath{\Dbar^0}\xspace}
\def\DzDzb   {\ensuremath{\Dz {\kern -0.16em \Dzb}}\xspace}
\def\Dp      {\ensuremath{D^+}\xspace}
\def\Dm      {\ensuremath{D^-}\xspace}

\def\DpDm    {\ensuremath{\Dp {\kern -0.16em \Dm}}\xspace}

\def\Bbar    {\kern 0.18em\overline{\kern -0.18em B}{}\xspace}

\def\Bz      {\ensuremath{B^0}\xspace}
\def\Bzb     {\ensuremath{\Bbar^0}\xspace}
\def\BzBzb   {\ensuremath{\Bz {\kern -0.16em \Bzb}}\xspace}
\def\Bu      {\ensuremath{B^+}\xspace}
\def\Bub     {\ensuremath{B^-}\xspace}

\def\BpBm    {\ensuremath{\Bu {\kern -0.16em \Bub}}\xspace}

\def\BorBbar    {\kern 0.18em\optbar{\kern -0.18em B}{}\xspace}
\def\DorDbar    {\kern 0.18em\optbar{\kern -0.18em D}{}\xspace}
\def\KorKbar    {\kern 0.18em\optbar{\kern -0.18em K}{}\xspace}

\mathchardef\Upsilon="7107
\def\Y#1S{\ensuremath{\Upsilon{(#1S)}}\xspace}

\mathchardef\Deltares="7101
\mathchardef\Xi="7104
\mathchardef\Lambda="7103
\mathchardef\Sigma="7106
\mathchardef\Omega="710A

\def\Deltabar{\kern 0.25em\overline{\kern -0.25em \Deltares}{}\xspace}
\def\Lbar{\kern 0.2em\overline{\kern -0.2em\Lambda\kern 0.05em}\kern-0.05em{}\xspace}
\def\Sigbar{\kern 0.2em\overline{\kern -0.2em \Sigma}{}\xspace}
\def\Xibar{\kern 0.2em\overline{\kern -0.2em \Xi}{}\xspace}
\def\Obar{\kern 0.2em\overline{\kern -0.2em \Omega}{}\xspace}
\def\Nbar{\kern 0.2em\overline{\kern -0.2em N}{}\xspace}
\def\Xb{\kern 0.2em\overline{\kern -0.2em X}{}\xspace}

\newcommand{\tev}{\ensuremath{\mathrm{\,Te\kern -0.1em V}}\xspace}
\newcommand{\gev}{\ensuremath{\mathrm{\,Ge\kern -0.1em V}}\xspace}
\newcommand{\mev}{\ensuremath{\mathrm{\,Me\kern -0.1em V}}\xspace}
\newcommand{\kev}{\ensuremath{\mathrm{\,ke\kern -0.1em V}}\xspace}
\newcommand{\ev}{\ensuremath{\mathrm{\,e\kern -0.1em V}}\xspace}
\newcommand{\gevc}{\ensuremath{{\mathrm{\,Ge\kern -0.1em V\!/}c}}\xspace}
\newcommand{\mevc}{\ensuremath{{\mathrm{\,Me\kern -0.1em V\!/}c}}\xspace}
\newcommand{\gevcc}{\ensuremath{{\mathrm{\,Ge\kern -0.1em V\!/}c^2}}\xspace}
\newcommand{\mevcc}{\ensuremath{{\mathrm{\,Me\kern -0.1em V\!/}c^2}}\xspace}
\newcommand{\kevcc}{\ensuremath{{\mathrm{\,ke\kern -0.1em V\!/}c^2}}\xspace}

\def\mus  {\ensuremath{\rm \,\mus}\xspace}

\def\ms         {\ensuremath{{\rm \,ms}}\xspace}     
\def\mus        {\ensuremath{\,\mu{\rm s}}\xspace}    

\def\to                 {\ensuremath{\rightarrow}\xspace}

\def\pep2{PEP-II}

\def\gsim{{~\raise.15em\hbox{$>$}\kern-.85em
          \lower.35em\hbox{$\sim$}~}\xspace}
\def\lsim{{~\raise.15em\hbox{$<$}\kern-.85em
          \lower.35em\hbox{$\sim$}~}\xspace}

\def\CP                {\ensuremath{C\!P}\xspace}

\def\jetset74   {\mbox{\tt Jetset \hspace{-0.5em}7.\hspace{-0.2em}4}\xspace}

\begin{document}

\title{Measurement of the CKM angle $\gamma$ at \babar\ }

\author{Emmanuel Latour, on behalf of the \babar\ Collaboration}

\address{CNRS-IN2P3, LLR Ecole Polytechnique, Route de Saclay, 91128 Palaiseau, FRANCE}

\ead{latour@poly.in2p3.fr}

\begin{abstract}
We present a short review of the measurements of the CKM angle $\gamma$ performed by the \babar\ experiment. 
We focus on methods using charged $B$ decays, which give a direct access to $\gamma$ and provide 
the best constraints so far. 
\end{abstract}

\section{Introduction}
The angle $\gamma\equiv \arg[-V_{ud}V_{ub}^*/V_{cd}V_{cb}^*]$ being one of the least precisely known parameter of the $CKM$ unitarity 
triangle (UT), much effort is being devoted to improve the precision of its measurement, which would allow to reach more stringent limits on the allowed region of the  global CKM fit of the UT \cite{Charles:2004jd}. 
Several methods involving charged $B$ decays have been used. They rely on the same principles (section \ref{sec1}), and only differ in  the
final state considered. We do not discuss the indirect measurements with neutral $B$ decays throught the measurement of $\sin(2\beta+\gamma)$. A more detailed review is available in \cite{Tisserand:2007qj}.

\section{Principle of the direct measurements of $\gamma$ with charged $B$ decays}\label{sec1}
These methods  all rely on $B^\pm \to D^{(*)}K^{(*)\pm}$ decays. The $D^{(*)}$ meson being an admixture of $D^{(*)0}$ and $\bar D^{(*)0}$, 
a decay to a final state accessible to both $D^{(*)0}$ and $\bar D^{(*)0}$ can proceed either through a $b\to c$ Cabibbo and color 
favored transition or a $b\to u$ Cabibbo and color suppressed one. The interference that takes place between these two 
amplitudes, respectively   $A(b\to c)\propto \lambda^3$ and 
$A(b\to u)\propto \lambda^3\sqrt{\bar \rho^2+\bar\eta^2}e^{i(\delta_B-\gamma)}$, give rise
to observables sensitive to $\gamma$. \\
Depending on the final state considered for the $D$, three methods have been proposed: the GLW, ADS, and GGSZ methods. They all 
present the advantage of being theoretically clean, i.e. free of penguin pollution. Furthermore they have the same three 
observables in common:  the strong and weak phase difference $\delta_B$ and $\gamma$, 
and the amplitude ratio $r_B\equiv \frac{|A(b\to u)|}{|A(b\to c)|}$. 
This feature enables to combine the different methods and to obtain better constraints on $\gamma$.\\
To each $B$ decay mode  $DK$, $D^*K$ and $DK^*$ corresponds a set of parameters noted $(r_B,\delta_B)$, $(r_B^*,\delta_B^*)$, $(r_{sB},\delta_{sB})$ respectively. These parameters are measured experimentally through
\CP asymmetries $\mathcal A$ and ratios of branching fractions $\mathcal R$:

\begin{eqnarray}
\mathcal A & \equiv & \frac{\Gamma(B^-\to D^{(*)}K^{(*)-})-\Gamma(B^+\to \bar D^{(*)}K^{(*)+})}{\Gamma(B^-\to D^{(*)}K^{(*)-})+\Gamma(B^+\to \bar D^{(*)}K^{(*)+})}\\
\mathcal R & \equiv & \frac{\Gamma(B^-\to D^{(*)}K^{(*)-})+\Gamma(B^+\to \bar D^{(*)}K^{(*)+})}{\Gamma(B^-\to D^{(*)0}K^{(*)-})+\Gamma(B^+\to \bar D^{(*)0}K^{(*)+})}
\end{eqnarray}

\section{The GLW method}
In the GLW method \cite{Gronau:1990ra,Gronau:1991dp}, the $D$ is reconstructed in a \CP eigenstate. \babar\ measurements \cite{Aubert:2005rw,Aubert:2004hu,Aubert:2005cc}  use both  $K^+K^-$, $\pi^+\pi^-$ \CP even modes, and  (except for the $D^*K$ analysis) the three 
$\KS\pi^0$, $\KS\phi$, $\KS\omega$ \CP odd modes.  The $D^*$ is
reconstructed in $D^*\to D^0\pi^0$, and the $K^*(892)$ in $K^{*-}\to \KS\pi^-$. 
$\mathcal A_{\CP\pm}$ is found compatible with 0 and $\mathcal R_{\CP\pm}$ with 1, so that $\gamma$ cannot be constrained with the GLW method alone. 
However, from these results one can  derive the cartesian coordinates $x_{\pm}\equiv r_B\cos(\delta_B\pm\gamma)$ and
the parameter $r_B^2=(\mathcal R_{\CP+}+\mathcal R_{\CP-})/2$. 
For the $DK^-$ channel,    \babar\ obtains $x_+=-0.082\pm0.052\pm0.018$, $x_-=0.102\pm0.062\pm0.022$ and $r_B^2=-0.12\pm0.08\pm0.03$.     
For the $DK^{*-}$ mode, \babar\ obtains $x_{s+}=-0.32\pm0.18\pm0.07$, $x_{s-}=0.33\pm0.16\pm0.06$ and $r_{sB}^2=0.30\pm0.25$.     
The precision obtained is already competitive with the one of the GGSZ method (section \ref{sec:ggsz}), so that the GLW method is useful in the measurement of $\gamma$ when combining the different methods.

\section{The ADS method} 
In the ADS method \cite{Atwood:1996ci,Atwood:2000ck}, the $D$ meson originating from the $b\to c$ transition is reconstructed in the doubly Cabibbo suppressed $K^+\pi^-$ mode, 
whereas the $\bar D$ coming from the $b\to u$ transition decays to $K^+\pi^-$ through a Cabibbo favored diagram. 
Despite the small branching fractions of these decays ($\mathcal O(10^{-6})$), their relative amplitude is comparable, hence the
 interference is large. Since the $D$ is not reconstructed in a \CP eigenstate, the decay amplitudes contain two  additional
 observables $r_D\equiv\frac{A(D\to K^+\pi^-)}{A(D\to K^-\pi^+)}$, which has been measured precisely ($r_D^2=0.376\pm0.009$), and 
$\delta_D$ the difference of strong phase  between the two amplitudes, which is unknown. \\
None of the $DK$, $D^*K$ and $DK^*$ channels have yielded any signal \cite{Aubert:2005pj,Aubert:2005cr}. 
No \CP asymmetries could be computed. Only upper
 limits were set for $D^{(*)}K$ modes ($R_{ADS}<0.029$, $R_{ADS}^{*(D^0\pi^0)}<0.023$, $R_{ADS}^{*(D^0\gamma)}<0.045$, $r_B<0.23$, $r_B^*<0.16$). The $DK^*$ channel gave $R_{sADS}=0.046\pm0.032$ and $r_{sB}=0.20\pm0.14$.\\
More recently, \babar\ also performed the ADS analysis of the $B^-\to[K^+\pi^-\pi^0]_DK^-$ decay \cite{Aubert:2006ga}, which has higher 
branching fractions and a smaller $r_D=(0.214\pm0.011)\%$, so that the sensitivity to $r_B$ is increased.  The additional complication in this
case comes from the strong variation of $A_D$ and $\delta_D$ in the Dalitz plane. This channel did not yield any signal, and only upper limits  $R_{ADS}<0.039$ and  $r_B<0.19$ (at 95$\%$ C.L.) were set.

\section{The GGSZ method}\label{sec:ggsz}

The GGSZ method \cite{Giri:2003ty} consists in studying three-body decays of the $D$, and perform a Dalitz analysis to account for the 
different intermediate states involved in the total amplitude.  The $B^-\to [\KS\pi^+\pi^-]_D K^-$ decays provide the best 
constraints on $\gamma$ so far. To parameterize the decay amplitude over the Dalitz plane ($m_{+}^2$,$m_{-}^2$), 
\babar\ uses the Breit-Wigner model:
\begin{equation}
\mathcal A(m_-^2,m_+^2)=\sum_{r}^{}a_re^{i\phi_r}\mathcal A_r(m_-^2,m_+^2)+a_{NR}e^{i\phi_{NR}}\label{eqBWM}
\end{equation}
which consists in a sum of 16 resonances and a non resonant term. 
The amplitudes and phases of Eq. (\ref{eqBWM}) are fitted on high purity flavor tagged $D^{*+}\to D^0(\KS\pi^+\pi^-)\pi^+$ control samples.
The fit of the $B^-\to [\KS\pi^+\pi^-]_D K^-$Dalitz plot is performed using the unbiased and gaussian cartesian coordinates $x_{\pm}$ and $y_{\pm}$. 
From these \CP parameters, a frequentist approach based on $n$-D Neyman confidence regions is used to obtain $r_B$, $\delta_B$ and $\gamma$.
\babar\ obtains $r_B<0.142$ and $r_B^*\in[0.016-0.206]$\cite{Aubert:2006am,Aubert:2005yj}. Combining results from $DK$ and $D^*K$ channels  gives
$\gamma[\rm{mod}\hspace{0.1cm}\pi]=(92\pm41\pm11\pm12)^{\circ}$. The $DK^*$ channel does not provide any constraint on $\gamma$, and only 
a loose  upper limit $\kappa r_{sB}<0.5$ is set.\\
The GGSZ method has also been used to study the $B^- \to [\pi^+\pi^-\pi^0]_DK^-$ decays \cite{Aubert:2007ii}. The analysis is similar to the $\KS\pi^+\pi^-$ analysis, except that
the Dalitz structure is different (it uses 15 resonances) and backgrounds are larger. Even the use of the cartesian coordinates led to non linear correlations, so the
fit is performed using the polar coordinates $\rho_{\pm}\equiv \sqrt{(x_\pm-x_0)^2+y_\pm^2}$ and $\theta_\pm\equiv \arctan\left( \frac{y_\pm}{x_\pm-x_0}\right)$, 
where $x_0=0.85$ is a change of variable constant. The results obtained on $\theta_+=(147\pm23\pm13)$ and $\theta_-=(173\pm42\pm19)$ are not precise enough to determine $\gamma$.
The error obtained on the values of $\rho_+=0.75\pm0.11\pm0.06$ and $\rho_-=0.72\pm0.11\pm0.06$ are small enough to be useful. However not attempt was done to combine these results 
with the other GGSZ analyses.

\section{Conclusion}

The measurement of the unitarity triangle angle $\gamma$ is very difficult with the current statistics. Among the three
 methods proposed using charged $B$ decays, the GGSZ analysis of $B^-\to [\KS\pi+\pi-]_D K^-$ decays provides the best contraints on $\gamma$ so far.
The GLW method  gives competitive errors on $x_\pm$, which is useful when combining the methods. Using neutral $B$ decays to have an indirect measurement of $\gamma$ 
helps in tightening the constraint on $\gamma$, however these methods suffer from the need of theoretical imputs and statistical limitation. 
The constraint obtained on $\gamma$  by combining all these methods and using both \babar\ and Belle results is $\gamma=(78^{+19}_{-26})^{\circ}$, still far from 
the global CKM fit not including these measurements, which yields $\gamma=(61.5\pm8.7)^{\circ}$.   

\begin{figure}[hb]
\begin{center}
\includegraphics[width=14pc]{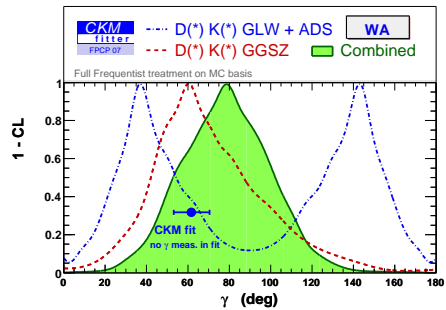}
\end{center}
\caption{\label{label}Global constraint on $\gamma$ using charged $B$ decays and combining \babar\ and Belle results.}
\end{figure}


\end{document}